\documentclass[runningheads]{llncs}
\usepackage{graphicx}
\usepackage{multirow}
\usepackage{comment}
\usepackage{float}
\usepackage{subfig}
\usepackage{color}
\usepackage{amsmath}
\usepackage{booktabs}
\usepackage{amsfonts}
\usepackage{booktabs}

\begin{document}
\title{EchoCP: An Echocardiography Dataset in Contrast Transthoracic Echocardiography for Patent Foramen Ovale Diagnosis}
\titlerunning{EchoCP}

%
\author{
Tianchen Wang\inst{1} \and 
Zhihe Li\inst{2} \and
Meiping Huang\inst{2} \and
Jian Zhuang\inst{2} \and
Shanshan Bi\inst{3} \and
Jiawei Zhang\inst{4} \and
Yiyu Shi\inst{1} \and
Hongwen Fei\inst{2} \and
Xiaowei Xu\inst{2}
}
\authorrunning{
T.Wang, et al.
}
\institute{University of Notre Dame, \\
\email{\{twang9, yshi4\}@nd.edu}
\and Guangdong Provincial People's Hospital\\
\email{
alphabet2018@163.com, huangmeiping@126.com, zhuangjian5413@tom.com, xiao.wei.xu@foxmail.com,}
\and Missouri University of Science and Technology\\
\email{bishanshan666@gmail.com}
\and Fudan University \\
\email{17110240008@fudan.edu.cn}
}
\maketitle               
\begin{abstract}
Patent foramen ovale (PFO) is a potential separation between the septum, primum and septum secundum located in the anterosuperior portion of the atrial septum. 
PFO is one of the main factors causing cryptogenic stroke which is the fifth leading cause of death in the United States.
For PFO diagnosis, contrast transthoracic echocardiography (cTTE) is preferred as being a more robust method compared with others. 
However, the current PFO diagnosis through cTTE is extremely slow as it is proceeded manually by sonographers on echocardiography videos. 
Currently there is no publicly available dataset for this important topic in the community.
In this paper, we present EchoCP, as the first echocardiography dataset in cTTE targeting PFO diagnosis. 
EchoCP consists of 30 patients with both rest and Valsalva maneuver videos which covers various PFO grades. 
We further establish an automated baseline method for PFO diagnosis based on the state-of-the-art cardiac chamber segmentation technique, which achieves 0.89 average mean Dice score, but only 0.60/0.67 mean accuracies for PFO diagnosis, leaving large room for improvement. 
We hope that the challenging EchoCP dataset can stimulate further research and lead to innovative and generic solutions that would have an impact in multiple domains. 
Our dataset is released \cite{echocp}.
\end{abstract}

\section{Introduction}


The foramen ovale is a physiological channel of the atrial septum and is usually functionally closed at birth.
Patent foramen ovale (PFO) arises when children older than 3 years do not have their foramen ovale closed, while approximately 20\%–25\% of adults have a PFO in the general population \cite{hagen1984incidence}.
PFO is one of the main factors causing cryptogenic stroke \cite{torti2004risk,khessali2012effect}, which is the fifth leading cause of death in the United States.
Recent studies show that the transcatheter closure of PFO reduces the recurrence of stroke at higher rates compared with medical therapy \cite{mas2017patent,saver2017long}.
Thus, it is essential to diagnose PFO in a fast and accurate way.

Various clinical methods have been used to diagnose PFO, such as contrast transcranial doppler echocardiography (cTCD), transesophageal echocardiography (TEE), and contrast transthoracic echocardiography (cTTE).
Although TEE is considered as the silver bullet for PFO diagnosis, patients find it difficult to successfully complete the Valsalva maneuver (VM) during TEE's invasive examination, which leads to a lower detection rate of right-to-left shunt (RLS) \cite{lip2014patent,del1998migraine}.
A large number of false negatives due to invalid VM during TEE examination further raise the need for other inspection methods.
As noninvasive methods are more acceptable for patients, both cTCD and cTTE are used to predict RLS by observing the number of microbubbles in the cranial circulation at the resting state and after VM.
However cTCD is not preferred as about 5\% of detected shunts do not correspond with PFO, which leads to lower sensitivity (68\%) and specificity (65\%) \cite{dao2011pfo,faggiano2012low}.
On the other hand, cTTE can isolate the source of RLS with a specificity of 97\% albeit with a slightly lower sensitivity of about 63\%\cite{truong2008prevalence}, which makes cTTE a simple, safe and reliable base as either the main PFO diagnosis method or a supplement for other methods. 


When using cTTE for PFO diagnosis, the apical four‐chamber view is generally selected which includes right/left atrium (RA/LA), and right/left ventricle (RV/LV) \cite{wang2019msu}. 
The presence of RLS is confirmed when microbubbles are observed in LV/LA within the first three cardiac cycles after contrast appearance in the right atrium during normal respiration or the VM. 
To perform an accurate and fast PFO diagnosis, there are several challenges that need to be addressed.
First, the existing approach is extremely time-consuming because the sonographers need to manually perform microbubble estimation as well as the PFO grade diagnosis on echocardiography videos.
Second, the existing approach is expensive. The representation of PFO in cTTE is easily confused with other cardiac diseases such as pulmonary arteriovenous fistula (PAVF), which requires experienced sonographers to diagnose accurately. 
Third, the data quality of cTTE is limited. In addition to the low resolution of echocardiography videos (e.g. $112\times112$\cite{ouyang2020video}), the heavy noise in images, the ambiguity at the components' boundaries, and the expensive data annotation further limit the quality of available data.
Meanwhile, the microbubbles to be expected in the cardiac chambers complicate the scenes. 
The automatic cTTE based PFO diagnosis still remains a missing piece due to the complexity of diagnosis and the lack of a dataset.


To the best of our knowledge, there are only three related echocardiographic datasets, however, all focus on other tasks with different annotation protocols, which is not applicable to PFO diagnosis through cTTE.
\textbf{CETUS} \cite{bernard2015standardized} is the first echocardiographic dataset introduced in MICCAI 2014. 
The main challenge is the LV endocardium identification, which contains 45 patients sequences with the annotations only contouring the LV chamber on end-diastolic (ED) and end-systolic (ES) instants.
\textbf{CAMUS} \cite{leclerc2019deep} was introduced to enhance the delineation of the cardiac structures from 2D echocardiographic images.
The datasets contains 500 patients with manual expert annotations for the LV, LA, and the myocardium regions.
\textbf{EchoNet-Dynamic} \cite{ouyang2020video} was introduced as a large echocardiography video dataset to study cardiac function evaluation and medical decision making.
The dataset contains 10,271 apical four-chamber echocardiography videos with tracings and labels of LV chambers on ED and ES instants.

In this paper we present EchoCP, the first dataset for cTTE based PFO diagnosis.
EchoCP contains both VM and rest echocardiography videos captured from 30 patients.
Data annotation including diagnosis annotation and segmentation annotation are performed by four experienced cardiovascular sonographers.
As there are more than a thousand images in each patient's video, sparse labeling (only select representative frames) of the segmentation is adopted \cite{wang2020ica}.
Other than existing echocardiography datasets, EchoCP is the first dataset targeting cTTE based PFO diagnosis.
In addition to the dataset, we also introduce a multi-step PFO diagnosis method based on a state-of-the-art chamber segmentation algorithm. 
Results show that the baseline method can achieve a 0.89 mean Dice score in segmentation and mean PFO diagnosis accuracies of 0.70/0.67 for VM and rest videos. 
As such, this dataset still calls for breakthroughs in segmentation/diagnosis algorithms for further improvement.

\begin{table}[t]
\centering
\caption{PFO diagnosis grade based on RLS through cTTE \cite{zhao2019modified}.}
\label{tab:pfo-rls}
\begin{tabular}{@{}lllll@{}}
\toprule
Grade & & RLS State & & Description \\ \midrule
0 & & None & & No microbubble observed in LV per frame. \\
1 & & Minor & & 1 to 10 microbubble observed in LV per frame.  \\
2 & & Moderate & & 11 to 30 microbubble observed in LV per frame.  \\
3 & & Severe & & More than 30 microbubble observed in LV per frame. \\ \bottomrule
\end{tabular}
\end{table}

\begin{table}[t]
\caption{Characteristics of the EchoCP dataset.}
\centering
\begin{tabular}{lll}
\toprule
Disease ($\#$PFO(\%) / $\#$Normal(\%)) & & 20(66.7\%) / 10(33.3\%)     \\
Sex ($\#$Male(\%) / $\#$Female(\%)) & & 17(57\%) / 13(43\%)     \\
Age (Mean$\pm$SD)  &  & 35.0$\pm$17.6       \\
Manufacturer     &  &  Philip EPIQ 7C (Probe S5-1, 1-5MHz) \\
Heart beat rate (Mean$\pm$SD)   & & 73.9$\pm$11.6       \\
Systolic/Diastolic blood pressure(mmHg)  &  & 120$\pm$11.6 / 74.3$\pm$10.8  \\
Patient's height ($cm$) & &  160.1$\pm$8.5              \\
Patient's weight ($cm$) & &  59.2$\pm$13.3              \\
Spatial Size of 2D images (pixels)  &  & 640$\times$480 \\
Video length (frames) &  & 954$\pm$232   \\ \bottomrule
\end{tabular}
\label{tab:dataset}
\end{table}

\begin{figure}[t]
    \centering
    \includegraphics[width=0.9\linewidth]{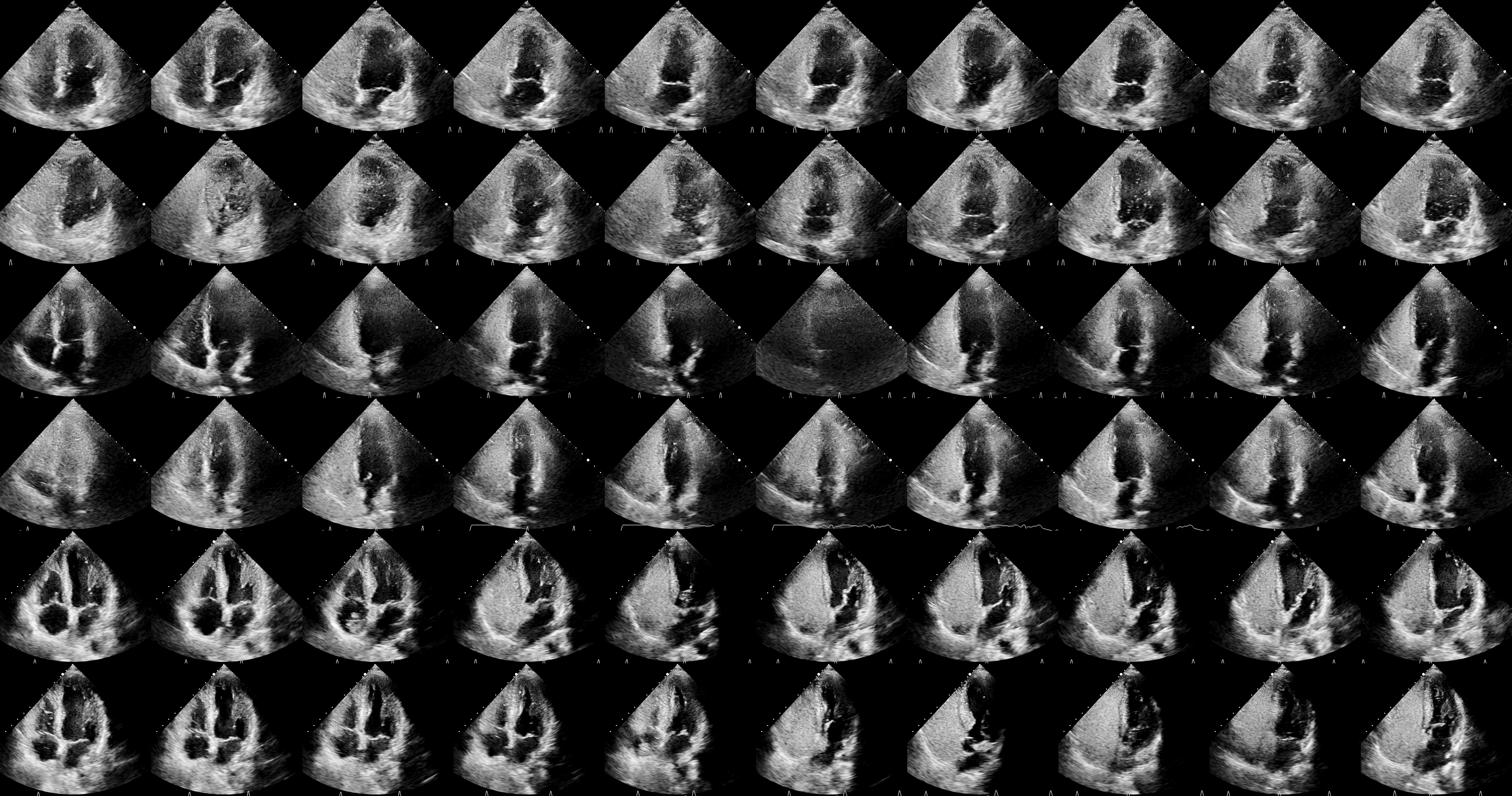}
    \caption{Representative frames from EchoCP. The ECG data, text labels, and ultrasound acquisition information are removed in pre-processing. The rest and VM states from three patients (top three rows and bottom three rows, respectively) are sampled.}
    \label{fig:frames}
\end{figure}

\section{The EchoCP Dataset}
\subsection{Data characteristics}
\subsubsection{Data acquisition}
Our EchoCP is made of echocardiography videos in cTTE for PFO diagnosis.
EchoCP contains cTTE videos from 30 patients. For each patient, two videos corresponding to the rest and VM state of the patients are captured. 
Note that in the rest state, patients just relax and breathe normally. While in the VM, patients need to close their mouths, pinch their noses shut while expelling air out as if blowing up a balloon.
The video is captured in the apical-4-chamber view and contains at least ten cardiac cycles.
For the VM state, the action is performed three to five times during acquisition, and we selected the most representative one. 
More characteristics of our EchoCP dataset are shown in Table~\ref{tab:dataset} and the representative frames are shown in Fig.~\ref{fig:frames}.
\subsubsection{Data annotation}
The data annotation of EchoCP includes two aspects: diagnosis annotation and segmentation annotation.
For diagnosis annotation, patients' reports are extracted from our medical system, and the diagnosis is further evaluated and confirmed by two sonographers.
For segmentation annotation, two sonographers are involved, and each annotation is annotated by one and evaluated by the other \cite{xu2020imagechd}.
In segmentation annotation, the sonographers manually draw the regions of RV/LV/RA/LA, which do not include myocardium and valve regions.
Due to the long echocardiography videos and the similarities among the frames, we select only representative frames for annotation. 
Frames in several representative situations where: 
(a) no microbubbles in all chambers, 
(b) many microbubbles in only RA, 
(c) many microbubbles in only RA and RV, 
(d) many microbubbles in all chambers, 
(e) moderate microbubbles in LA and LV, and
(f) a few microbubbles in RA and RV are selected.
For each selected situation, four frames are annotated (two in the diastole and two in the systole). 
Due to the variety of patients, the videos from several patients may not cover all the above representative situations.
Thus, usually 20 to 24 frames are annotated for each patient.

\subsection{PFO diagnosis and evaluation protocol}
\label{sec:criterion}
As mentioned above, PFO diagnosis through cTTE relies on the states of microbubble in LV after the microbubble solution injection.
Specifically, during the test, a mixture of saline and room air is agitated to present as a good contrast agent, which is then vertically injected into the left cubital vein as a bolus both at rest and during VM.
After a few cardiac cycles, the microbubbles would enter RV and then RA regions, and the two chambers would be full of microbubbles.
We then observed the number of microbubbles shown in the LV region within three cardiac cycles.
If the microbubbles are not shown in LA and LV chambers, the patient may have a negative PFO (grade 0).
Otherwise, we proceed with the diagnosis by classifying RLS states based on the number of observed microbubbles, as shown in Table~\ref{tab:pfo-rls}.

It is noted that for PFO diagnosis through cTTE, the decision on VM videos plays the dominant role.
However, we still proceed with diagnosis method on both VM and rest videos because the result from the rest videos can act as a reference.
Meanwhile for cardiac function diagnosis on other diseases such as PAVF, the RLS states classification in the rest videos would carry more weight in the diagnosis \cite{sagin1998contrast}.
Therefore in the following proposed method, we process on both VM and rest videos for a more complete study.

\begin{figure}[t]
    \centering
    \includegraphics[width=1.0\linewidth]{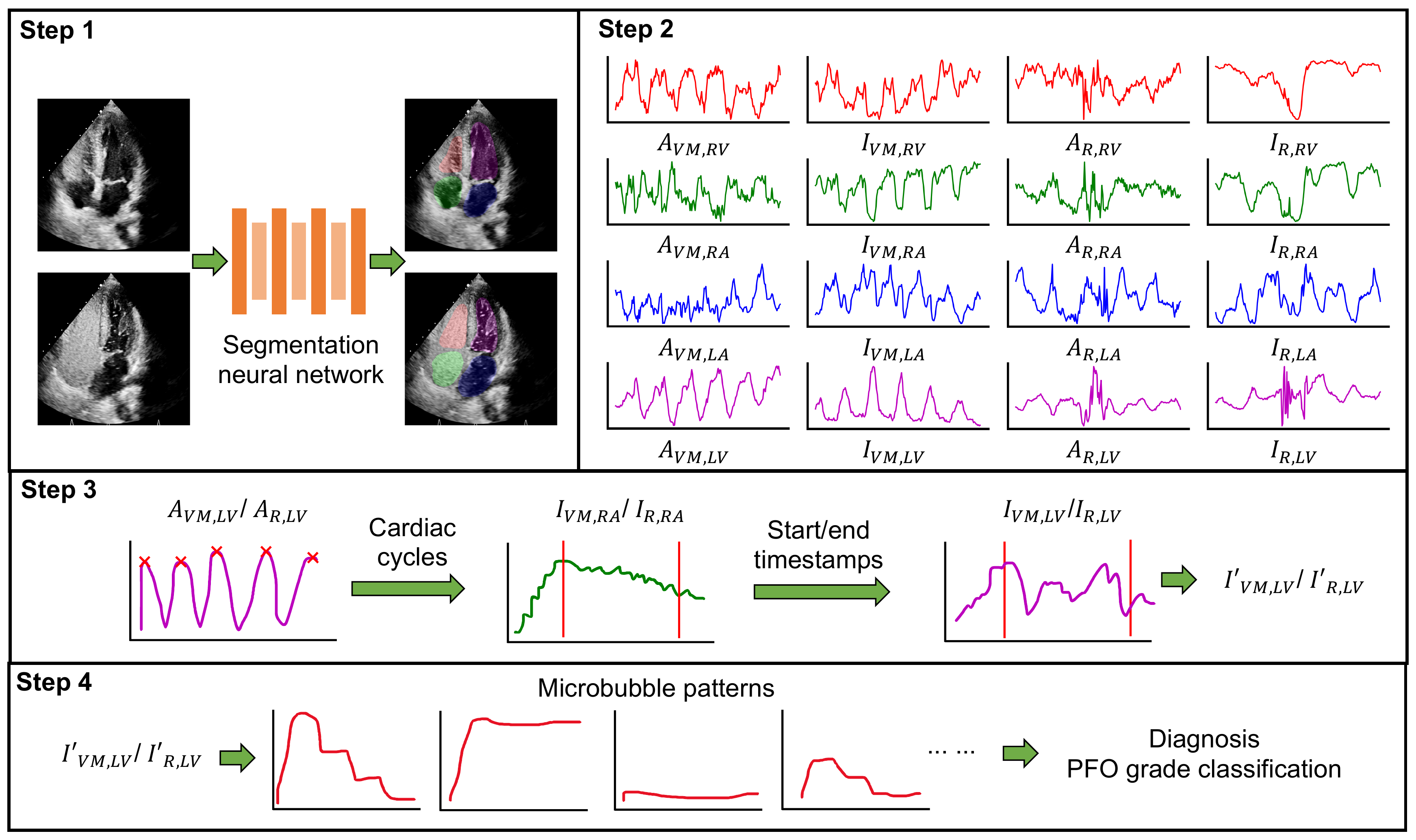}
    \caption{The proposed baseline method has four steps: (1) segment to get the regions of four chambers through videos; (2) calculate the area ($A$) and the mean intensity ($I$) in rest ($R$) and VM ($V\!M$) states of four chambers (RV, RA, LA, LV); (3) get the start timestamp (RA full of microbubbles) and the end timestamp (three cardiac cycles), cropping $I_{VM,LV}$ and $I_{R,LV}$ to get $I'_{VM,LV}$ and $I'_{R,LV}$; (4) Fit $I'_{VM,LV}$ and $I'_{R,LV}$ to microbubble patterns for diagnosis.
    }
    \label{fig:method}
\end{figure}

\section{Experiments of Baseline Method}
\label{sec:baseline}
In this section, we introduce a baseline method for PFO diagnosis using EchoCP.
The method is based on the criterion in Section~\ref{sec:criterion} and a state-of-the-art segmentation framework.
Note that although our goal is to perform video-based PFO diagnosis which is basically a video classification task, we cannot apply an end-to-end video classification method to it.
The first reason is that the lengths of captured videos are long with a large variety, ranging from 537 to 1386. 
It is not computationally feasible to propagate all frames at the same time for classification. 
The second reason is that the PFO diagnosis labels are video-level rather than frame-level, which means that we do not know the PFO grade of videos until the whole videos are seen and analyzed. 
We cannot classify or label a random small clip. 
Therefore, a multi-step method based on diagnosis criterion and a state-of-the-art framework is an applicable approach.
As shown in Fig.~\ref{fig:method}, the proposed method has four steps as illustrated below.

\begin{itemize}
\item {\bf Step 1}. We use an encoder-decoder style neural network to segment all the frames in both VM and rest videos, and obtain the regions of four chambers (RV, RA, LV, LA) as the output.
As our task is to segment four chambers that are not available in previous echocardiography datasets \cite{bernard2015standardized,leclerc2019deep,ouyang2020video}, we trained the model only with the labeled frames in EchoCP. 
We use a representative state-of-the-art 2D Dynamic U-net (DynUnet) \cite{isensee2019automated} to proceed segmentation task with the resized frames (384$\times$384).
For data augmentation, since the number of available annotations is limited, we do not proceed with a heavy augmentation such as distortion and affine transformation due to the relatively fixed positions of target chambers in the echocardiography videos.
Thus, we choose a light augmentation that only consists of a random Gaussian noise added to images.
The learning rate is set to 1e-3, the optimizer is set to Adam \cite{kingma2014adam}, and the loss is Dice loss \cite{milletari2016v}. 
\item {\bf Step 2}. We further process the obtained segmentation maps of four chambers in VM and rest videos.
For each chamber (RV, RA, LA, LV) in each video ($V\!M$, $R$), two values through all the frames are calculated: area ($A$), and the mean intensity ($I$).
Correspondingly, a total of 16 values are obtained.
Note that the curves shown in Fig.~\ref{fig:method} are cropped for visualization. 
\item {\bf Step 3}. We first obtain the cardiac cycles, which can be estimated by the ED/ES cycles of LV.
This can be achieved by locating the maximum area of LV during the cardiac cycles, thus finding the peaks of $A_{LV}$.
Then we locate the timestamp when RA is filled with microbubbles as the start.
This can be estimated as the first time when $I_{RA}$ is larger than an intensity threshold.
Then we crop the $I_{LV}$ between the start timestamp and the three cardiac cycles after that as $I'_{LV}$ for the next step.
\item {\bf Step 4}. We apply the $I'_{LV}$ to fit the predefined microbubble patterns.
Along with the maximum intensity in $I'_{LV}$, we proceed with the PFO diagnosis by classifying the grade based on the predefined intensity thresholds.
\end{itemize}
Note that we randomly choose the 30 videos from 15 patients for training process, including both the Step 1 segmentation and the Step 4 diagnosis thresholds set up.
We use another 30 videos for testing.


\begin{figure}[t]
    \centering
    \includegraphics[width=1.0\linewidth]{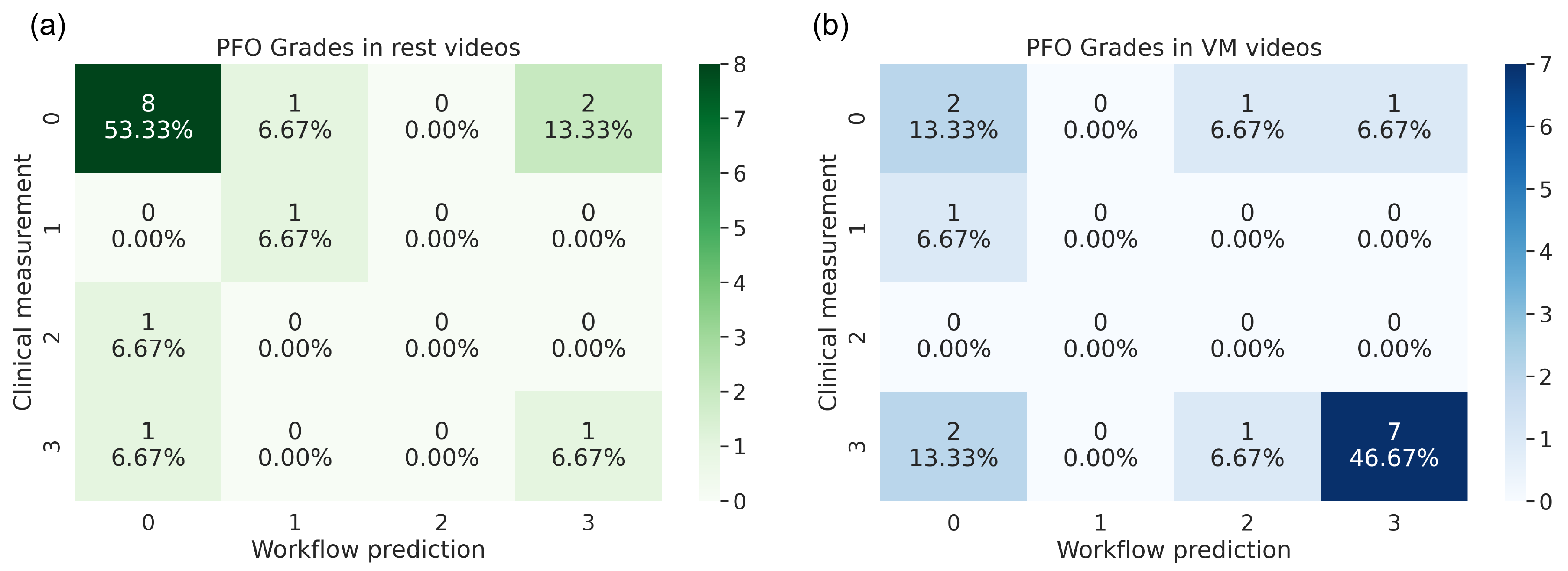}
    \caption{Confusion matrices of the proposed method for PFO diagnosis on VM videos (a) and rest videos (b), built on a state-of-the-art segmentation method.}
    \label{fig:result}
\end{figure}

\begin{figure}[t]
    \centering
    \includegraphics[width=1.0\linewidth]{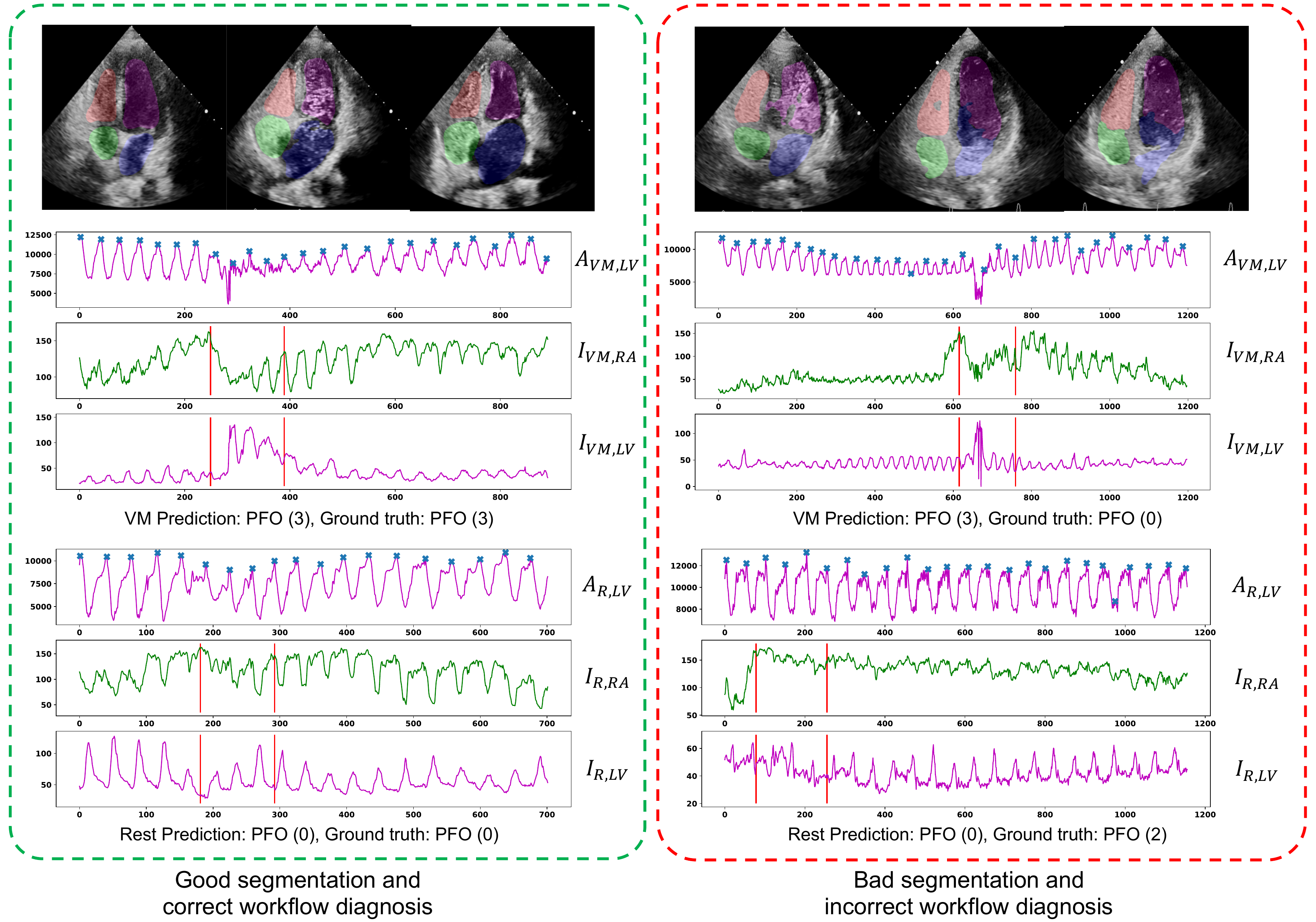}
    \caption{Examples of good/bad segmentation results, and correct/incorrect diagnosis.}
    \label{fig:bad_result}
\end{figure}



\section{Results and Analysis}
Following the proposed baseline method, we obtained $A_{VM,LV}$, $I_{VM,RA}$, $I_{VM,LV}$ and $A_{R,LV}$, $I_{R,RA}$, $I_{R,LV}$ for each patient.
The four chambers segmentation task in Step (1) has achieved a mean Dice score of 0.89.
The confusion matrices between the PFO prediction results and the human clinical measurements (as ground truth) on both VM and rest videos are shown in Fig.~\ref{fig:result}.
The labels align the PFO grades in Table~\ref{tab:pfo-rls}.

In PFO diagnosis with VM videos, the method achieves an average accuracy of 0.60. with precision/recall of 0.88/0.70, 0.0/0.0, 0.0/0.0, and 0.40/0.50 on PFO grade 3/2/1/0.
For rest videos, the method achieves an average accuracy of 0.67, with precision/recall of 0.33/0.50, 0.0/0.0, 0.50/1.00, and 0.80/0.73 on PFO grade 3/2/1/0.
The first major misdiagnosis is in VM videos where the grade-3 PFO cases are predicted as negative (indicated by a number of 4 in Fig.~\ref{fig:result}(a)).
The second major misdiagnosis is in rest videos where the negative cases are predicted as grade-3 PFO (indicated by a number of 3 in Fig.~\ref{fig:result}(b)). 

We further show the success and failure examples in Fig.~\ref{fig:bad_result}.
For segmentation, the good prediction examples would have the complete chamber regions and the correct contouring around and between the chambers' boundaries, which are not shown in bad prediction examples.
For PFO diagnosis, the good examples would have the accurate estimation of cardiac cycles, the starting timestamp as observation, and number of microbubbles for fitting patterns.
In the bad diagnosis examples of VM videos, the cardiac cycles are not accurately detected. 
Meanwhile due to the inaccurate segmentation, the chambers' regions are not correctly detected, which results in an intensity peak in LV, and the corresponding false negative.
For the bad diagnosis examples in rest videos, although the cardiac cycles are well detected and the segmentation performs well, the microbubble estimation through intensity in LV does not show the pattern, leading to a false positive.
This shows that the current proposed microbubble estimation approach by chamber intensity may not apply to all scenarios in echocardiography videos, 
where the periodical appearance of the myocardium background in target chambers accounts for more intensity changes compared with the filled microbubbles.

There are several workable approaches that could improve the diagnostic accuracy.
First, more accurate segmentation is appreciated, especially when microbubbles flow in the chambers. 
The current segmentation method we used is image-based that ignores the temporal correlation between the frames in videos.
A segmentation method with tracking function may work better.
Second, more accurate microbubble estimation is required.
Our current approach estimates the number of microbubbles by chamber's intensity. 
A more precise approach is to localize and count the microbubbles.
Third, few shot end-to-end video classification is an attractive overall solution. 
By learning the limited data and extracting extremely compact representations from the long videos, the end-to-end diagnosis approaches would be possible, which requires deep collaboration between artificial intelligence experts and radiologists.

\section{Conclusion}
We introduce to the community EchoCP \cite{echocp} as the first echocardiography dataset in cTTE for PFO diagnosis, in hopes of encouraging new research into the unique, difficult, and meaningful topic. 
Based on this dataset, we also introduce a baseline method for PFO diagnosis which consists of four chambers segmentation, data extraction, and pattern matching. 
The baseline method achieves a Dice score of 0.89 on the segmentation and the diagnosis mean accuracies of 0.60 on VM videos and 0.67 on rest videos, leaving considerable room for improvement.

\bibliographystyle{splncs04}
\bibliography{paper798}

\end{document}